\documentclass[twocolumn,prb,aps,a4paper,amsmath,amssymb,showpacs,
superscriptaddress,floatfix]{revtex4}
\usepackage{color}
\usepackage{graphicx}

\usepackage[hypertex]{hyperref}

\renewcommand{\Re}{\mathop\mathrm{Re}\nolimits}
\renewcommand{\Im}{\mathop\mathrm{Im}\nolimits}

\def\be{\begin{equation}}
\def\ee{\end{equation}}
\def\bea{\begin{eqnarray}}
\def\eea{\end{eqnarray}}

\begin{document}

\title{Fractional Quantum Hall Effect and vortex lattices. II}

\author{S.V. Iordanski}

\affiliation{Landau Institute for Theoretical Physics, Russian Academy of Sciences,Kosygin str.2 Moscow, 117334 Russia}
\author{D. S.Lubshin}
\affiliation{Landau Institute for Theoretical Physics, Russian Academy of Sciences,Kosygin str.2 Moscow, 117334 Russia}
\date{\today}

\begin{abstract}
It is demonstrated that all observed fractions at moderate Landau level fillings
for the quantum Hall effect can be obtained without recourse to the
phenomenological concept of composite fermions. The possibility to have
the special topologically nontrivial many-electron wave functions is
considered. Their group classification indicates the special values of
of electron density in the ground states separated by
a gap from excited states.  These gaps were calculated for some lattices in
a simplified model.
\end{abstract}

\maketitle

The experimental discovery of Integer Quantum Hall Effect (IQHE) by
K.v Klitzing (1980) and Fractional Quantum Hall Effect (FQHE) by Tsui, Stormer
and Gossard (1982) was one of the most outstanding achievements in condensed
matter physics of the last century.

Despite the fact that more than twenty
years have elapsed since the
experimental discovery of quantum Hall Effect (QHE), the theory of
this phenomenon is far from being complete (see reviews \cite{qh},\cite{nqh}).
This is primarily true for the Fractional Quantum Hall Effect (FQHE),
which necessitates the electron--electron interaction and can not
 be explained by the one-particle theory, in contrast to the
IQHE. The most successful variational many-electron wave
function for explaining the 1/3 and other odd inverse fillings was constructed
by Laughlin(\cite{lgh1},\cite{lgh2}). The explanation of other observed fractions was obtained by
various phenomenological hierarchial schemes with construction of the "daughter"
states from the basic ones (Haldane 1983,Laughlin 1984, B.Halperin 1984).

 In those works, the approximation of extremely high magnetic
field was used and all states were constructed from the states at
the lowest Landau level. However, this does not conform to the
experimental situation, where the cyclotron energy is of the order of
the mean energy of electron--electron interaction. Moreover, this
approach encounters difficulties in generalizing to the other
fractions. Computer simulations also give a rather crude approximation
for the realistic multiparticle functions, because the
number of particles in the corresponding calculations on modern
computers does not exceed several tens.

The most successful phenomenological description is given by the
Jain's model of "composite" fermions \cite{j1},\cite{j2}, which predicts the
majority of observed fractions. According to this model, electrons
are dressed by magnetic-flux quanta with magnetic field
concentrated in an infinitely narrow region around each electron. It
is assumed that even number of flux quanta provides that these
particles are fermions. The inclusion of this additional
magnetic field in the formalized theory leads to the so-called Chern--
Simons Hamiltonian. This approach is described in details in \cite{h1}.

However, this theory gives an artificial 6-fermionic interaction whereas

the actulal calculations use quite crude mean field approximation of the
"effective" magnetic field as the sum of the external magnetic field and some
additional artificial  field that provides the total magnetic flux quanta
 in accordance with Jain's model of composite fermions.

In the present work we shall show  how to remove some restrictions of
 Jain-Chern-Simons model and obtain a more
general and more simple model which does not change the  standard Coulomb
interaction of electrons. The main
concept is associated with the notion of topological classification of
 quantum states. There is a number of
topological textures in condensed matter physics: Vortex lattices in a rotating
superfluid, Abrikosov vortices in
superconductors, skyrmions in 2d electron systems at integer fillings of
Landau levels. It is difficult to give an exact
topological classification of the multiparticle wave function for various
physical systems. Possibly the most simple
and general definition can be done using canonical transformation of
the field operators of the second quantization.
 The canonical transformation of the field operators is one which does not
 change their commutation relations.
 We do not consider the statistical transmutations which possibly can
not be achieved at low energies
considered in condensed matter physics. In general there must be the proper
topological classification of  the canonical transformations itself.

In this work we consider the simplest case of the fermion canonical
 transformation not including spin degrees
of
freedom and assuming the full polarization of 2d electrons
\begin{gather}
\psi({\bf r})=e^{i\alpha({\bf r})}\chi, 
\\
\psi^{+}({\bf r})=\chi^{+}
e^{-i\alpha({\bf r})},
\end{gather}
with  $\alpha({\bf r}$ ) having vortex kind singularities. It is evident that
 $\chi$ and $\chi^{+}$ satisfy Fermi kind
commutation relations if $\psi$ and $\psi^{+}$ satisfy them. Inserting these
 expressions into the  standard
hamiltonian for the interacting  electrons (with omitted spin indices )

\begin{multline}
\label{ham}
H=\frac{\hbar^2}{2m}\int \psi^{+}(-i{\bf \nabla }-\frac{e}{c\hbar}{\bf A})^2
\psi d^2r+\\
\int \frac{U({\bf r-r'})}{2}\psi^{+}({\bf r})\psi^{+}({\bf r'})
\psi({\bf r'})\psi({\bf r})d^2rd^2r',
\end{multline}
we get  a new Hamiltonian
\begin{multline}
\label{nham}
H=\frac{\hbar^2}{2m}\int \chi^{+}(-i{\bf \nabla}+ {\bf \nabla}\alpha -
\frac{e}{c\hbar}{\bf A})^2\chi d^2r
\\
+\int \frac{U(|{\bf r-r'}|)}{2}\chi^{+}({\bf r})\chi^{+}({\bf r'})
\chi({\bf r'})\chi({\bf r})d^2rd^2r',
\end{multline}
where $U(r)$ is Coulomb interaction. We want to consider a set of periodic
vortexlike singularities in
${\bf \nabla}\alpha$. Vector ${\bf \nabla}\alpha$ can be expressed in terms of
 Weierstrass zeta function
 used in
the theory of the rotating superfluids \cite{tk} given by the converging series
\begin{equation}
\label{z}
\zeta=\frac{1}{z}+\sum_{T_{nn'}\ne 0}(\frac{1}{z-T_{nn'}}+\frac{1}{T_{nn'}}+
\frac{z}{T_{nn'}^2}),
\end{equation}
where $z=x+iy$ is a complex coordinate on 2d plain, $T_{nn'}=n\tau+n'\tau'$
and $\tau$, $\tau'$ are  the minimal
complex periods \cite{ui} of the vortex lattice. The phase factor $e^{i\alpha
}$ will be simple function on
2d plain if ${\bf \nabla}\alpha=K(\Re\zeta,\Im\zeta)$ and
\begin{equation}
\label{al}
\alpha({\bf r})=K
\int_{\bf r_0}^{\bf r}(\Re\zeta dx+\Im\zeta dy),
\end{equation}
with integer $K$ of any sign. The quantity $K$ and the periods $\tau$, $\tau'$
define the topological class of multiparticle
wave function. The transformed Hamiltonian (\ref{nham}) can not be restored to
the initial form (\ref{ham})
by any smooth finite transformation of the function $\alpha$ .
That
 makes it topologically stable.
We  shall investigate the pecularities of the ground state and excitations for
this model at low temperature.

 Having in mind large magnetic fields it is interesting to consider the
 simplified version of the hamiltonian
 (\ref{nham}) without the interaction term
 \begin{equation}
 \label{h'}
 H'=\frac{\hbar^2}{2m}\int\chi^{+}\left[-i{\bf \nabla}+{\bf \nabla}\alpha-
 \frac{e}{c\hbar}{\bf A(r)}\right]^2\chi d^2r.
 \end{equation}

 This Hamiltonian has  properties  close to the Hamiltonian with a
 constant magnetic field. Indeed the translation on any period ${\vec \tau}$ of
 the
 vortex lattice gives an additional constant in the brackets
 \begin{gather}
 \label{tr}
 {\bf r}\to{\bf r}+{\vec \tau},
 \\
 \begin{split}[-i{\bf \nabla}+{\bf \nabla}\alpha-\frac{e}{c\hbar}&A({\bf r})]\to
 \\
 [-i{\bf \nabla}+&{\bf \nabla}\alpha-\frac{e}{c\hbar}{\bf A}({\bf r})+
 {\vec \delta}({\vec \tau})-\frac{e}{c \hbar}{\bf A}(\vec \tau)],
 \end{split}
 \end{gather}
 due to the properties of Weierstrass function
 $\zeta(z+ \tau)=\zeta(z)+\delta(\tau)$ and the linear dependence of the
 external vector potential ${\bf A}({\bf r})$ at constant magnetic field. The
 additonal constant terms can be removed by the gauge transformation of the
 field operators $\chi$, $\chi^{+}$. Thus the proper magnetic translation does
 not change Hamiltonian (\ref{h'}).

 If we introduce the ``effective'' vector potential
 ${\bf A}_{\mathrm{eff}}={\bf A}-\frac{c\hbar}{e}{\bf  \nabla}\alpha$, the magnetic
 translation is
  given
 by the transformation
 \begin{equation}
 \label{mt}
 T_m({\vec \tau})\chi=\chi({\bf r}+{\vec\tau}) \exp\left(\frac {ie}{c\hbar}
 {\bf A}_{\mathrm{eff}}({\vec \tau}){\bf r}\right),
 \end{equation}
 for any real period of the vortex lattice.

  It is easy to connect ${\bf A}_{\mathrm{eff}}({\vec \tau})$ with the ``effective''
  magnetic flux through the unit cell of the vortex lattice given by the
  contour
  along it's boundaries
 \begin{gather}
  \Phi=\oint{\bf A}_{\mathrm{eff}}d{\bf r}= {\bf A}_{\mathrm{eff}}({\vec \tau_1})
  {\vec \tau_2}-
  {\bf A}_{\mathrm{eff}}({\vec \tau_2}){\vec \tau_1}.
 \end{gather}
  On the other hand it can be calculated directly using the definition of
  ${\bf A}_{\mathrm{eff}}$
  \begin{equation}
  \label{fi}
  \Phi={\bf B}{\vec \tau}_1\times {\vec \tau}_2+K\Phi_0,
  \end{equation}
  where $\Phi_0=2\pi\frac{e}{c\hbar}$ is the quantum of the flux, $B$ is
  the external magnetic field.

  As was shown by E.Brown (1964) \cite{br}, J.Zak (1964) \cite{zk} (see also
  \cite{lp9}) 
  the simple finite representation of the ray group of magnetic translations
  can be obtained only for rational number of the flux quanta per unit cell,
  \begin{equation}
  \label{phi'}
  \Phi=\frac{l}{N}\Phi_0=Bs+K\Phi_0,
  \end{equation}
  where $s$ is the area of the unit cell of the vortex lattice, $l$ and $N$
  are integers without common factors.

  Thus  the situation for the vortex lattices  is isomorphous to
  the case of uniform magnetic field with  a rational number of the flux quanta
  per the unit cell. Therefore it is possible to use all the argumentation
  following the paper \cite{br} in constructing of the finite representation
  for the ray group of magnetic translations. In order to construct the finite
  representation one must impose certain boundary conditions on the solutions
  of Schroedinger equation with the hamiltonian (\ref{h'}). The simplest is
  the magnetic periodicity,
  \begin{equation}
  \label{mp}
  T_m({\bf L})\,\chi({\bf r})=\chi({\bf r}),
  \end{equation}
  where ${\bf L=L_1,L_2}$ define the size of the sample,
   ${\bf L_1}=NM_1{\vec \tau_1}$,${\bf L_2}=NM_2{\vec \tau_2}$ with integer
   $M_1,M_2$. It easy to show that any magnetically translated   function
   $\chi$ according to (\ref{mt}) will also satisfy (\ref{mp}). The simplest
   realization is the vortex lattice  consisting of exactly $N\times N$ unit
   cells.

   This conditions is the analog of Born-von Karman conditions in the absence
   of magnetic field. Indeed in a large enough system the density of states
   practically does not depend on the exact form of boundary conditions. But
   the restriction to the finite representations is important.

   The matrices of the representation are
   \begin{gather}
   \notag
   D_{jk}(0)=\delta_{jk},
   \\\label{rp1}
   D_{jk}({\vec \tau_1})=\delta_{jk}\exp{i(j-1)\frac{l}{N}},
   \\\notag
   D_{j,k}({\vec \tau_2})=\delta_{j,k-1} \,(\rm{mod}\,  N)(j,k=1,2...N),
   \end{gather}
   and the general matrix of the representation
   \begin{multline}
   \label{rp2}
   D_{jk}(n_1{\vec \tau_1}+n_2{\vec \tau_2})=
   \\
   \exp\left\{i\pi\frac{ln_1}{N}[n_2+2(j-1)]
   \delta_{j,k-n_2}\right\}(\rm{mod}\,  N).
   \end{multline}

   The traces of all matrices are zero except identity which has a trace equal 
   to
   $N$. The sum of the squares of traces is $N^2$. Therefore the representation
   is irreducible. The square of the dimensionality is also $N^2$ therefore
   there can be no other nonequivalent representation. The dimensionality of
   the representation  gives  $N$ fold degeneracy of the energy levels for
   Hamiltonian (\ref{h'}). The number of the equivalent representations in a
   regular representation is also $N$.

   These equivalent representations depend on the choice of the gauge.  We choose the total effective vector potential
   giving nonzero flux through the unit cell of the vortex lattice with $ A_{\mathrm{eff},y}=0 $ assuming $ A_{\mathrm{eff},x
   }\neq 0$,
   and the basic periods for the vortex lattice $\vec{ \tau}_1=(\tau_x,0)$ and $\vec{\tau}_2$  with both components.
   Let us take a magnetic cell  containing  $N$ cells in ${\vec{\tau}}_2$ direction and only one
   cell in the $\vec{\tau}_1$ direction. The states of the crystal of $N\times N$ cells can be labelled by  the
   quasimomentum ${\bf q}= q_1{\bf b}_1+q_2{\bf b}_2 $ where
   \begin{gather}
   {\bf b_2}=\frac{2\pi \hat{z}\times \vec{\tau}_1}{|\vec{\tau}_1\times\vec{\tau}_2|},
   \\
   {\bf b_1}=\frac{2\pi \vec{\tau}_2 \times \hat{z}}{N|\vec{\tau}_1\times\vec{\tau}_2|},
   \end{gather}
      are the basic vectors of the reciprocal lattice. There is no translation in $\vec{\tau}_2$ direction therefore we can
   put $q_2=0$ but there are $N-1$ translations in $\vec{\tau}_1$ direction. Because $N$ translations in that
   directions must give the phase $N\vec{\tau}_1{\bf b}_1q_1=2\pi r$  we must put $q_1=r/N$ with
   $r=0,...,N-1$ . These equivalent representations together with $N$ times degeneracy due to the dimension
   of the representation give the regular representation of the translation group in $N\times N$ cells crystal.

   The limitation to the single magnetic cell $N{\vec \tau}_1,N{\vec \tau}_2$
   can be easy removed by the consideration of the vortex lattices with
   dimensions $N_1{\vec \tau}_1,N_2{\vec \tau}_2$ where $N_1=NM_1$, $N_2=NM_2$
   for integer $M_1$,$M_2$. The representations of the larger group of
   $N_1\times N_2$ operations can be formed from the already discussed.

   For this
   group there are $M_1M_2$ representations of dimensionality $N$. The matrices
   corresponding to the translations ${\vec \tau_1}$,${\vec \tau_2}$ differ from the
   already given only by a phase factor. These representations can be labelled
   by a vector with reciprocal space components of ${\ q}_1$, ${ q}_2 $
   \begin{equation}
   \label{rp3}
   D^{\bf q}({\vec \tau_j})\equiv\exp(-i{ q_j\tau_j})D({\vec \tau_j}), 
   \end{equation}
   where $j=1,2$ and possible values of ${\bf q}_j$ are given by
   \begin{gather}
   \label{rp4}
   q_j=\frac{2\pi C_j}{N_j\tau_j},\qquad j=1,2
   \\
   C_1=0,1,...M_1-1,
   \\
   C_2=0,1,...M_2-1.
   \end{gather}
   In  general  every vector ${\bf q}$  corresponds to some irreducible representation of the translation group.
   The     possible
   domain of ${\bf q}$  is defined by the Brilluin zone for the given periodical part of magnetic field.
   For the same gauge  with $A_{\mathrm{eff},y}=0$ the domain for $q_1$  will be unchanged by the constant part
   of
   the effective magnetic field with a nonzero flux.
   Therefore it is $M_1N$  different values of the  $q_1$ . But the domain for $q_2$  can be
   reduced if one introduces a new unit cell extended in ${\vec{\tau}}_2$ direction  with
   ${\bf A}_2= N{\vec{\tau}}_2$ in order to have the integer number of the flux
   quanta . That reduced $q_2$ domain is $1/N$  smaller then in the basical reciprocal lattice. The total number of
   the different values  for ${\bf q}$ for the irreducible representations inside the reduced Brilluin zone will be
   $M_1M_2N$. It must be multiplied by $N$ due to the dimensionality of the representation giving $M_1M_2N^2$
   various states equal to the number of states in the primary Brilluin zone without  constant part of
   the effective magnetic field. That calculation  is in a close analogy to the case of zero flux periodical magnetic field
   where the irreducible representations are abelian one dimensional representations of the translation group instead
   of the discussed $N$  dimensional irreducible representation.
   Because vectors ${\bf q}$ are quasicontinuous inside the reduced Brilluin zone there is no
   energetical gaps inside this set of the states. If one takes only a part  of this set there will be no energetical gap
   at the transitions to the rest empty states.  We suggest that
     the set  of $M_1M_2N^2$ states is divided by some
   gap from the analogous set with the higher energies like it was for the case of a zero flux periodical; magnetic field.
  That must be checked numerically.

      At large magnetic fields the Hamiltonian (\ref{h'}) will be dominating in the
   full Hamiltonian (\ref{nham})
   because it linearly depends on magnetic field while the interaction term
   is proportional to the square root of it. In this case the energy of the
   ground state including the interaction can be obtained by the perturbation
   theory
   \begin{multline}
   \label{gs}
   E_0=E_0'+\\
   \frac{1}{2}\int U_c(|{\bf r-r'}|)
   \langle\chi^{+}({\bf r})\chi^{+}({\bf r'})\chi({\bf r'})\chi({\bf r})\rangle d^2rd^2r',
   \end{multline}
   here $E_0'$ is the energy for the lowest set of the states giving the irreducible represemtations  and
   the angle brackets denote the
   average  over the Slater determinant of the wave functions  of the set.
  ( fully filled ground state   of
   the Hamiltonian
   (\ref{h'})). The energy gap dividing the ground state from
   the next set of states with the higher energies at large magnetic
   fields must be proportional to the value of the external magnetic field.
   In the
   performed experiments \cite{dlg1} the linear dependence of the jump for
   electron chemical potential in a strong magnetic fields  was
   observed for the fractions 1/3 and 2/3.
   The  expression for the gap must be obtained by the numerical calculation
   of  Bloch functions
    for the given representation and is dependent on $K$,
   $N$,$l$ and periods $\tau_i$.

   One can see that in the model of the vortex lattices  the gap does not depend
   exclusively on the interaction term like it was suggested in most of
   theoretical works based on the degeneracy of the ground Landau level.
   Opposite , it is almost independent from the
   interaction in strong magnetic fields. The resolution of this paradox is
   the same  as in the
   rotating superfluid. The origin of the observed vortex lattices in a rotating
   superfluid is connected with the thermodynamic energy in the rotating frame
   $E'=E-{\bf \Omega M}$,where ${\bf \Omega}$ is the angular velocity
    and ${\bf M}$ is the angle momentum of the superfluid. That requires
    the superfluid velocity to be equal to the velocity of the solid body
    rotation
    and the vortex lattice is a good approximation in a superfluid. Really it is
    connected with a different dependence of the energy on the size of the
    system giving the preference to the solid body rotation irrespective to the
    microscopical internal structure of the superfluid.

The previous group analysis valid for a rational number of the flux quanta
show that the energy gaps are opened at the special electron densities
corresponding to one electron per each unit cell of the vortex lattice, that
gives according to Eq.(\ref{phi'}) the electron
density
\begin{equation}
\label{ed}
 n_e=\frac{B}{\Phi_0}\frac{N}{l-NK}
 \end{equation}
 and must correspond to the filled set of bands
 obtained from $\frac{S}{s}$ states in the absence of the average magnetic
  field.
 Here, $S$ is the sample area. Simple analysis states \cite{lp9} that
 this initial band is split into $q$ subbands, each being (odd $q$) $q$-
 fold or (even $q$) $q/2$-fold degenerate, and with the fraction of the
 number of states in each subband being (odd $q$)equal to $1/q^2$ or (even $q$)
 to $2/q^2$. However, the total number of states in all subbands is
 $S/s$. One can assume that, even in the presence of
 interaction, these states are separated from the higher-energy states
 by the greatest gap.  Note that the
 evenness of the $K$ number is immaterial, because
 the Fermi commutation rules
 or the operators $\chi$ and $\chi^{+}$ are fulfilled automatically and
 have no relation to the topological number $K$ unlike Jain-Chern-Simons theory.
The occurrence of any
specific numbers of vortex flux quanta can be dictated
by the ground-state energy . Indeed the experimental electron density is
defined by the gate voltages or the density of the compensating charge but the
observation of the specific Hall plato can be dictated by the value of the gap
at the given temperature and the purity of the sample.
The observed fractions in FQHE correspond to the
following tables:

\vspace{5mm}
$K=-2,\;\;l=1$

\vspace{3mm}

\begin{tabular}{|c|c|c|c|c|c|c|c|c|c|}
\hline
$N$ & 1&2&3&-5&-2&-3&-4&4&$\infty$\\
\hline
$\nu$
&$\frac{\mathstrut 1}{\mathstrut 3}$
&$\frac{2}{5}$
&$\frac{3}{7}$
&$\frac{5}{9}$
&$\frac{2}{3}$
&$\frac{3}{5}$
&$\frac{4}{7}$
&$\frac{4}{9}$
&$\frac{1}{2}$\\
\hline
\end{tabular}

\vspace{5mm}

That fractions correspond to celebrated Jain's rule \cite{j2}. Half filling of the
Landau level $n_e=\frac{B}{2\phi_0}$ in the external field corresponds to a
vanishingly small effective magnetic field (zero number of flux
quanta per unitary cell).

Other observed fractions correspond to
\vspace{5mm}

$K=-1,\;\;l=1$
\vspace{3mm}

\begin{tabular}{|c|c|c|c|}
\hline
$N$&-4&4&2\\
\hline
$\nu$&$\frac{\mathstrut 4}{\mathstrut 3}$&$\frac{4}{5}$&$\frac{2}{3}$\\
\hline
\end{tabular}

\vspace{5mm}
where one has double of the fraction 2/3, and

\vspace{5mm}
$K=-1,\;\;l=2$

\vspace{3mm}
\begin{tabular}{|c|c|c|c|c|}
\hline
$N$&-7&-5&5&2\\
\hline
$\nu$&$\frac{\mathstrut 7}{\mathstrut 5}$&$\frac{5}{3}$&$\frac{5}{7}$&$
\frac{1}{2}$\\
\hline
\end{tabular}

\vspace{5mm}
Here one has not observed double of the fraction 1/2 with the gap
($ B_{\mathrm{eff}}\neq0 $).
 In the following section we consider the related questions more thoroughly.

 \section{The energy gain due to the vortex formation}

 Previously we were acting quite formally introducing the vortex lattice and
 
 obtaining the transformed Hamiltonian by conjecture. In this section we show
 that the free energy of any state without a macroscopical current is unstable
 in the external magnetic field due to the formation of the isolated vortex.
 The change of the free energy of the charged system in the given external
 magnetic field is (see e.g.\cite{LL2})
 $$ \delta F=-\frac{1}{c} \int {\bf A}\delta{\bf j}dV
 $$
 where ${\bf A}$ is the external vector -potential ,$\delta{\bf j}$ is
  the
 change of the current and the integration is over the volume of the sample.
 That corresponds exactly to 2DES, where the external magnetic field can not be
 essentially changed by a weak 2D current. Thus this equation can be integrated
 to give
\begin{equation}
 \label{Fe}
  F=E-\frac{1}{c}\int {\bf A}{\bf j}d^2r
 \end{equation}
 Where $ E $ is the internal energy.

 The standard assumption in the theory of 2DES in a strong magnetic field
 was the possibility to construct the ground state by the projection on the
 states in the lowest Landau level(e.g. \cite{BIE}). In that case  the
 average electron current vanish on the distances of the order of magnetic length.

 It is possible to calculate the free energy change (\ref{Fe}) due to
 the formation of the isolated vortex. It is convenient to use the axial gauge
 where  the external vector-potential is
 $${\bf A }({\bf r})= \frac{1}{2} rB{\bf e}_{\phi}$$
 where ${\bf e}_{\phi}$ is the unit vector in the azimuthal direction. The
 effective vector-potential is the sum of the external-vector potential and the
 additional vector -potential of the vortex
 $$\delta {\bf A}=\frac{c\hbar}{e}\frac{K}{r}{\bf e}_{\phi}$$

  which is the change due to the formation of the vortex at the origin. The
  corresponding operator of the electrical current reads
  \begin{multline}
  \label{cur}
  \hat{\bf j}=\frac{\hbar
  e}{2Ml_B}\left\{\psi^{+}\left(-\frac{i\partial}
 {\rho\partial\phi}+\frac{\rho}{2}{\bf e}_{\phi}-\frac{K}{\rho}{\bf e}_{\phi}\right)\psi+\right.
 \\\left.
 \left(\frac{i\partial}{\rho\partial\phi}\psi^{+}+\frac{\rho}{2}{\bf e}_\phi\psi^{+}-
 \frac{K}{\rho}{\bf e}_{\phi}\psi^{+}\right)\psi\right\},
  \end{multline}
 where $\rho=r/l_B$. We shall measure all energies in the units  of $\hbar^2/(2Ml_B^2)$.
 
 We assume the lowest Ll partially filled and the projected form of the fermi
 operators is
 \begin{gather}
 \psi= \sum_{m}\exp({-im\phi})R_m(\rho)c_m,
 \\
 \psi^{+}=\sum_{m}\exp({im\phi})R_m(\rho)c_m^{+},
 \end{gather}
  where $c_m,c_m^{+}$ are fermi operators and
 $$R_m(\rho)=\rho^m\exp({-\rho^2/4}) N_m^{-1/2},$$
 where $N_m=2^m l_B^{2(m+1)}m!$. It is easy to show that the total azimuthal
 current through any ray $\phi=const$ vanish at $\delta{\bf A}=0$.
 The second term with the magnetic moment  in eq.(\ref{Fe}) reads
 $$ F_2=2\pi\int _0^R\rho d\rho\left[\rho\sum_{m>0}R_m^2\left(\frac{m}{\rho}-\frac{\rho}{2}-|\frac{K}{\rho}\right)
 \langle c_m^{+}c_m\rangle\right].$$
 The angular brackets denote quantum mechanical average.

 We consider  large distances from the positions of the  vortex where the perturbation of the basic state is small and
 this expression can be calculated up to the first order as the average over the supposed projected ground state.The
 first term in brackets is linear in the sample size and  can be neglected compare to the term due to the formation of
 the vortex proportional to the sample area
 \begin{equation}
 \label{mm}
 F_2=-K\int_0^R  n_e2\pi rdr,
 \end{equation}
 where $n_e$ is the average ectron density and the integral gives the total number of electrons in the sample.
 It is essential that the main contribution  comes from the large distances where the states are distorted quite weakly
 therefore the interaction and microscopical structure are not changed.

 The calculation of the internal energy can proceed along the same lines. The change of the internal energy due to
  vortex formation is given by the kinetic energy term which reads
  \begin{multline}
  \label{ie}
  E'=\int_0^R  \frac{2K}{\rho}\sum_{m>0}(\frac{\rho}{2}-\frac{m}{\rho})R_m^2\langle c^{+}_mc_m\rangle2\pi  rdr+
  \\
  \int_0^R  \frac{K^2}{\rho^2}\sum_{m>0}R_m^2\langle c^{+}_mc_m\rangle2\pi rdr.
  \end{multline}
  In the same order of the perturbation theory that gives  logarithmic dependence on the sample size. It is evident
  for the last term. The first term gives also the logarithmic contribution at large distances. The finite value of $E'$ is obtained by some cut in the vortex core on small distances where the electron density must be reduced.

 Thus the vortex formation gives the
 gain in the energy of the large enough samples when the negative  magnetic
 moment term in the free energy exceeds the logarithmic increase of the internal
 energy in eq( \ref{Fe}). We see that the supposed ground state projected on
 the lowest Ll is unstable to the vortex formation.
 This statement is independent  of the
 details of the microscopic structure or the interaction and is valid only due
 to the different size dependence of the internal energy and the magnetic moment term
 of 2DES in a close analogy to the case to the rotating liquid(\cite{lp9}).

 The regularization is essential to obtain this result. There are two possibilities for the regularization
 known from the theory of superfluid $^{3}He$  (\cite{V}). The simplest  are singular vortices with the hard core defined by
 Coulomb interaction and the  atomic structure of the underlying semiconductor of the heterostructure. That gives
 the estimate of the order of the atomic Bohr radius for the core size, The other possibility is a soft core  with the size defined by the
 extension of the phase space. In 2DES that is either electron spin or the isospin connected with the next level of
 the size quantisation for the electron motion in the perpendicular to 2D plain direction.  That corresponds to so called Skyrmion texture (\cite{S}) .That gives the core of the order of the magnetic length. That case is much more complicated and
  we restrict the  considertion by the lattices of hard core vortices.

 \section{The  half filling of  Ll}.
 The above considerations also give the thermodynamic preferences to the
 vortices with the minimal winding number $K=1$. The limiting case of the
 half filling of Ll can be obtained either from the larger densities when
 $N\to(-\infty$ )or from the smaller densities when $N\to(+\infty$). The unit
 cell of the vortex lattice contain two vortices with $K=1$ exactly compensating
 the flux of the external magnetic field. The effective magnetic field is
 periodic with the zero flux and it is possible to use the gauge giving also
 periodic effective vector-potential. We suppose that the vortex lattice has the
 form of of two parallel triangular lattice for each of vortices in the unit cell  displaced
 by the distance $({\vec{ \tau}'=( \vec{\tau}_1+\vec{\tau}_2})/3$ as shown on Fig.1.The Bravais lattice corresponds
 to the positions of one vortex in the unit cell.  The positions of the other
 vortex are shown by crosses.

 Because the total flux is zero one has the abelian translation group and the
 electron states can be classified by a quasimomentum. The ground state
 must correspond to the filling of the lowest band. The Brilluin zone  for the
 triangle lattice has the form of the hexagon with the primitive vectors of the
 inverse lattice
 \begin{gather}
 {\bf b_1}=\frac{2\pi}{s}({\vec{\tau}_2}\times\hat z ),
 \\
 {\bf b_2}=\frac{2\pi}{s}(\hat z\times{\vec{\tau}_1}),
 \end{gather}
  where $s$ is the area of the unit cell of the vortex lattice. There are
 two nonequivalent vectors on the face of the Brilluin zone
 ${\bf q_1}=(q_x,0)$ and the other obtained from the first by
 $\frac{2\pi}{6}$ rotation with $q_1=\frac{b}{\sqrt 3}$.as it is shown in fig.2.
  The whole space group  is
 isomorphous to the space group of the honeycomb 2D crystal like graphene but
 the Hamiltonian corresponds to the periodic vector-potential instead of the
 periodic potential
 \begin{gather}
 H'=\frac{(\hbar)^2}{2M_e}\int \psi^{+}\left[-i{ \nabla}_-\frac{e}{c}
 {\bf A}_{\mathrm{eff}} ({\bf r})\right]^2\psi d^2{\bf r}.
 \end{gather}

 The star for ${\bf q_1}$ consist of two rays ${\bf q_1,q_2}$ .The small
 representation of the space group corresponds to the rotations by
 $\pm\frac{2\pi}{3}$ giving two equivalent vectors $ {\bf q_1}+{\bf b}_i $ and
 the reflection $ y\to-y$ leaving ${\bf q}_1$ invariant and giving the same
 lattice after the nontrivial translation changing the positions of the crosses to
 the positions of the points in the Bravais lattice: $(r_y|{\bf \tau'})$.
  We use the standard notification (\cite{BP})  for the
 space group elements: $r_y$ is the reflection,$\tau_{r_y}$ is the corresponding
 translation . The representation of the space group can be obtained as some
 ray representation of the small group leaving ${\bf q_i}$ invariant
(including equivalent vectors)(see e.g.\cite{BP}) with the multiplication
 law
 $$D(r_1)D(r_2)=\omega(r_1,r_2)D(r_1r_2)$$
 for the matrices of the the space group representation. The representation
 coinside with the known representation for graphene having two dimensional
 representation at the points ${\bf q_1,q_2}$ on the faces of the Brilluin zone.

 Therefore we have no Fermi surface as usually suggested but two Fermi points
 at the electron density corresponding to the half filling of the Ll. This result corresponds to the symmetry
 between the electrons and holes.
  That gives the absence of energetical gap
 and conical Diracs spectrum
 $$\epsilon_i=\epsilon^0+|{\bf q-q_i}|v_F,$$
 in the vicinity of these points with $v_F\sim(\hbar)/(l_BM_e)$. The quantity
 $\epsilon^0$ gives the electron chemical potential at the half filling.

 The absence of the gap can also explains the interaction of 2D electrons with  acoustical phonons. Thus one can get the explanation of the attenuation of the surface waves
 in SAW experiments (\cite{w1})  like it
 was done by using Jain-Chern Simons theory.\cite{h1}

 \section{The numerical calculation of the gaps}

 As was shown in the introduction the irreducible representations of the translation group for any vortex lattice
 with the rational number of the flux quanta are given by eq.(\ref{rp3}) with any ${\bf q}$ from the reduced Brilluin
 zone with the specified gauge  ${A_{\mathrm{eff},y}=0}$. The representations with a different ${\bf q}$ are different.
 It means according to the general theorems (\cite{BP}) that the states corresponding to the different representations
 are orthogonal to each other. The dimensionality of the representation gives  $N$ degenerate states which are also orthogonal  because the matrices of the representation are unitarian. In order to perform the numerical calculations
 it is possible to use only one basic function of the representation (\ref{rp3}).

 The translation group is the subgroup of the space group for $2d$ vortex lattice.  The procedure to find the irreducible representations  of the space group is well elaborated  for the ordinary crystal without magnetic flux per unit cell. It
 is possible to have a generalization for the rational number of flux quanta. For the simplest ``simmorf'' case the
 space group is given by the product  of the translation subgroup and the subgroup of the point symmetry of the
 rotations and the reflections. It gives the possibility to have some additional degeneracy due to the subgroup of
 the point symmetry at the specific values of the wave vector ${\bf q}$. For the non ``simmorf'' space group there
 is a possibility to have the representation of the higher  dimensionality then $N$ at some specific values of ${\bf q}$.
 This additional gegeneracy is generated if the translation subgroup does not commute with all elements of the
 space group at these specific values of ${\bf q}$. For the nonspecific values of ${\bf q}$ the situation is the same
 as in the ``symmorf"
  case and one has a set of ${\bf q}$ with $N$ fold degenerate states. We consider in the
 calculations only the ``simmorf'' case.

 Using the known irreducible representation for the translation group in the presence of the magnetic field  with
 the rational flux  (\ref{rp3}) one can get the most simple partner function (\cite{br})
 \begin{multline}
 \label{pf}
 f_0^{\bf q}=\frac{1}{N}\sum_{n_1,n_2}\exp(i{\bf q}_1n_1{\vec{\tau}}_1)\exp(i{\bf q}_2n_2N{\vec{\tau}}_2)\times
 \\
 T_m(n_1{\vec{\tau}}_1)T_m(n_2N{\vec{\tau}}_2)g({\bf r}).
 \end{multline}
 Where the vectors in the reduced Brilluin zone are
 $${\bf q}_1=\frac{r_1}{NM_1}{\bf b}_1, r_1= 0,...,NM_1-1$$
 $${\bf q}_2=\frac{r_2}{M_2}{\bf b}_2, r_2=0,...,M_2-1$$
 and the summation is over $0\le n_1<NM_1$, $0\le n_2<M_2$. The other $N-1$ partner functions are
 given by the action of the other translations
 \begin{equation}
 \label{pf2}
 f_{m'}^{\bf q}=\exp(-im'{\bf q}_2{\vec{\tau}}_2)T_m(-m'{\vec{\tau}}_2)f_0^{\bf q}({\bf r})
 \end{equation}
 for $m'=1,...,N-1$. These functions has the same energy and are not essential  for the calculation of the energy
 $\epsilon({\bf q})$. It is possible to calculate  this energy by the minimization of the mean value of the Hamiltonian
(\ref{h'}) over the partner function (\ref{pf}) by the specifying of the unknown function $g({\bf r})$.That is the standard way to use group symmetry of the hamiltonian. But in this work we have used more universal  and powerfull method developed in (\cite{O}) based on the fast Fourrier transform method obtaining all basic functions of the representatiom
and the energies as well. As described this method is suitable for any shape of the periodic magnetic field.

We used the regularization of the periodic part of the "effective" vector potential  $ A'_{\mathrm{eff},x}({\bf r})$ by  the Fourrier truncation of the corresponding periodic magnetic field representing the proper  delta -function by
the finite sum

$ \delta({\bf r})\approx \sum_{p_x=0}^P\sum_{p_y=0}^P cos\frac{2\pi p_x x}{L}cos\frac{2\pi p_yy}{L}$
where $P=10$ and $L$ is the size of the unit cell of the vortex lattices.

 It is ruther cumbersome to perform the analysis for the various vortex lattices with the same $l/N$ flux quanta per unit cell. Instead we have tried several specific lattices and  calculated the energies for the ground zone and the next
 by the effective numerical method and check the existence or the absence of the energetical gap between.
 In this calculations we consider only the sinplified model with the Hamiltonian (\ref{h'}) neglecting the interaction
 term. We use  the definte gauge where $ A_{\mathrm{eff},y}=0$ and the  vortex lattice has periods
 ${\vec{\tau}}_1=( \tau_{1,x},\tau_{1,y}=0)$ and ${\tau}_2=(\tau_{2,x},\tau_{2,y})$. The Shroedinger equation corresponding to Hamiltonian (\ref{h'}) acquires the form
\begin{equation}
\label{hc}
\left(\frac{1}{2}( -i\partial_x +y+A'_x)^2+\frac{1}{2}(-i\partial_y)^2\right)\psi=\epsilon\psi
\end{equation}
Where ${\bf A'}$ is the periodic part of the effective vector potential with zero flux. Here the distances are measured
in the magnetic length  units for the constant effective magnetic field $B^0=\frac{l}{l-NK}B$  , and the energy is
in the units of the corresponding cyclotron energy $\omega_c^0=\omega_c(B)|\frac{l}{l-NK}|$. The solution
$\psi$ of the eq.(\ref{hc}) must satisfy the magnetic periodic conditions (\ref{mp}) on the sample boundaries.
 The preliminary
results of the numerical calculations are given by the following table, where
 by $K$ we denote the number of flux quanta carried by
a single vortex.  Square (S) or triangular (T)  lattice corresponds to the structure formed by vortices
themselves.  Energies are given in units of $\hbar eB/mc$, where $B$ is the external
 uniform magnetic field. We write $\epsilon_{min,i}$, $\epsilon_{max,i}$  for the minimal and the maximal
 values of the  energy $\epsilon({\bf q})$ in two lowest zones $i=1,2$, and $\Delta$ is the energetical gap between.

\begin{tabular}{|c|c|c|c|c|c|c|c|c|c|}
\hline
Lattice & l & K & N & $\nu$ & $\epsilon_{1min}$ & $\epsilon_{1max}$ & $\epsilon_{2min}$
& $\epsilon_{2max}$ &$\Delta$ \\ \hline
S & 1 & -2 &  1 & 1/3 & 0.263 & 0.353 & 0.444 & 0.491 & 0.0909\\ \hline
S & 1 & -2 &  2 & 2/5 & 0.269 & 0.293 & 0.392 & 0.405 & 0,0986\\ \hline
S & 1 & -2 &  3 & 3/7& 0.265 & 0.269 & 0.337 & 0.363 & 0.0677\\ \hline
S & 1 & -2 & -2 & 2/3 & 0.274 & 0.337 & 0.490 & 0.720 & 0.153\\ \hline
S & 1 & -1 &  2 & 2/3 & 0.245 & 0.277 & 0.525 & 0.594 & 0.248 \\ \hline
T & 1 & -2 &  1 & 1/3 & 0.249 & 0.305 & 0.434 & 0.580 & 0.128 \\ \hline
T & 1 & -2 &  2 & 2/5 & 0.227 & 0.234 & 0.360 & 0.384 & 0.125 \\ \hline
T & 1 & -2 & -2 & 2/3 & 0.278 & 0.373 & 0.411 & 0.657 & 0.037 \\ \hline

\end{tabular}\\

As an example in fig. 3 is shown the electron dispersion law for $l=1$, $K=-1$,  $N=2$.

\section{Conclusion}
Thus we have reproduced the key statement of the Jain's theory of composite
fermions  and obtained the explanation of practically all
observed fractions
at moderate Landau level fillings in an unified frame without any hierarchial schemes. The preliminary results were
published in(\cite{I})
The numerical calculations give the evident energetical gaps at the experimentally observed  electron densities
at FQHE. Our description of the states for 2DES at the FQHE conditions is a kind of the mean field approximation or
more exactly the method of the selfconsistent  effective vector-potentia. We did not consider the fluctuations of the
vortex field assuming zero temperature. It is well known that in 2d the thermal fluctuations destroy the periodic order
of the crystall. The same must be true for the vortex lattice. It is reasonable to suppose that nevertheless  the energetical gap for the charged exitations survive and the electrons form a special electron liquid which can have
some kind of the plastic flow in the absence of the crystallic order.  The observed Hall current may be the realization
of this flow in the presence of the external electric field. The Hall constant may be defined by the mean electron density in the domain of this flow like it is in IQHE. The domains of the electron localization on the existing in the sample impurities will induce Hall plateau because the electron density in the flow domain is unchanged if the electron chemical potential corresponds to the energy of the localized states. The plausibility of this  qualitative picture must be checked by more detailed investigations.

\section{Acknowledgments}
Authors express their gratitide to L.P. Pitaevsky, V.F. Gantmakher,
V.T. Dolgopolov, M.V. Feigelman and I.V. Kolokolov
for the usefull discussions of the various questions concerning the subject of this work. The work was supported by the
Program "Quantum Macrophysics "  of RAS Presidium, the grant RFBR and the grant by the President of RF for
the support of the scientific scools.

\newpage
\section{Graphics}

\begin{center}
\begin{figure*}[h]
\includegraphics[width=0.9\linewidth]{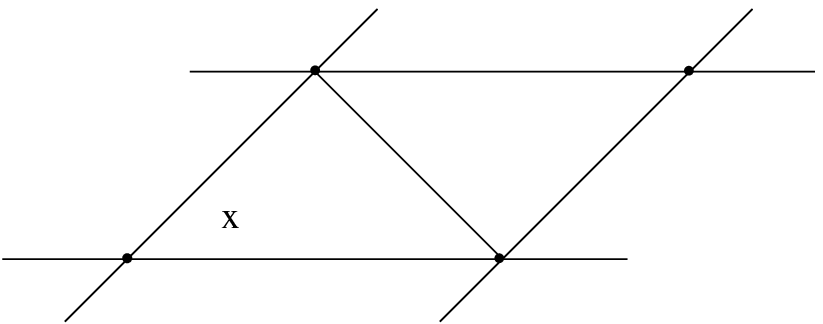}
\caption{The vortex lattice at half filling of the Ll.}
\label{fig:fig1}
\end{figure*}
\end{center}

\newpage
\begin{center}
\begin{figure*}[h]
\includegraphics[width=0.9\linewidth]{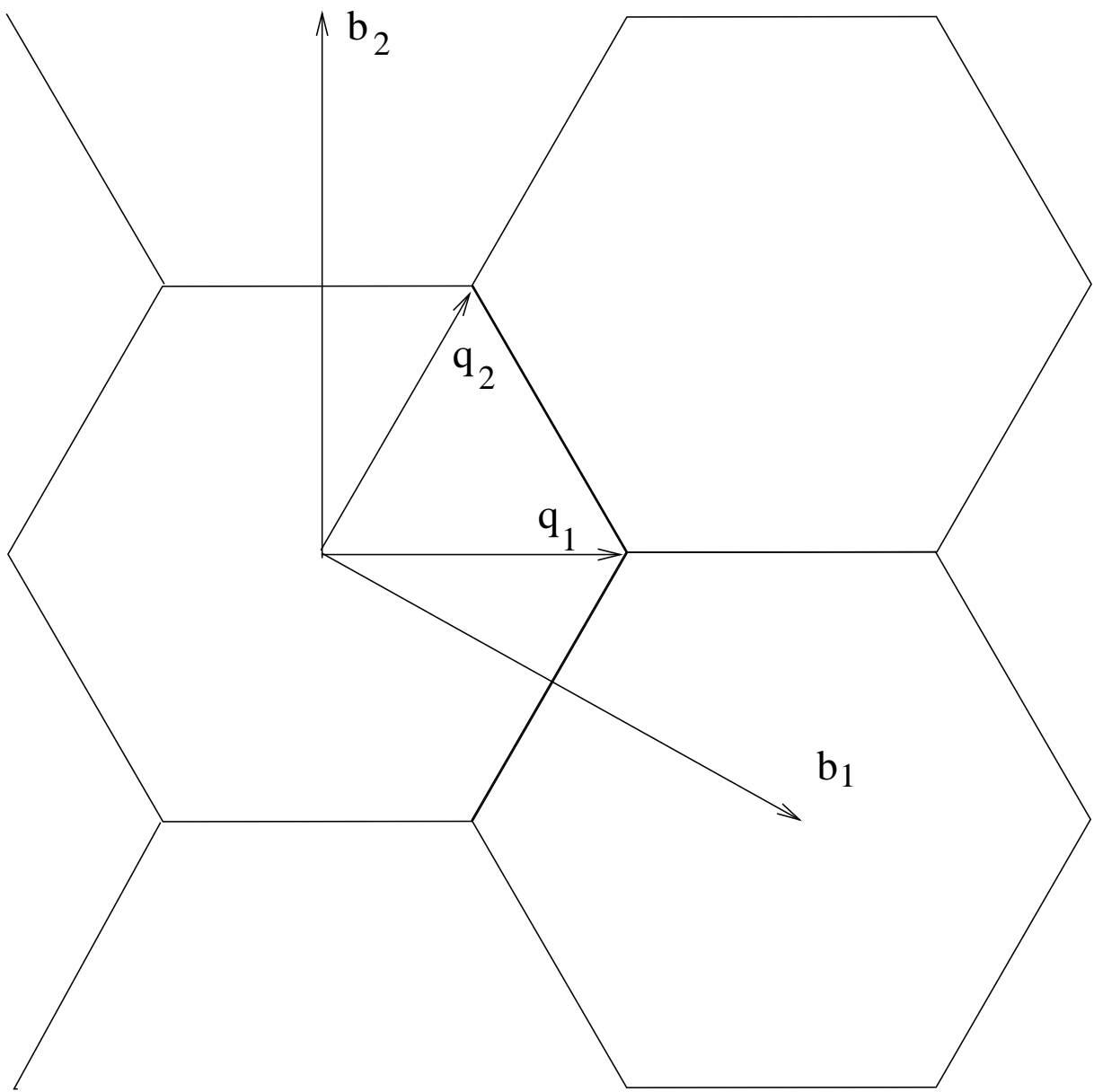}
\caption{The reciprocal lattice and Brilluin zone at  half filling of the Ll.}
\label{fig:fig2}
\end{figure*}
\end{center}

\newpage
\begin{center}
\begin{figure*}[h]
\includegraphics[width=0.9\linewidth]{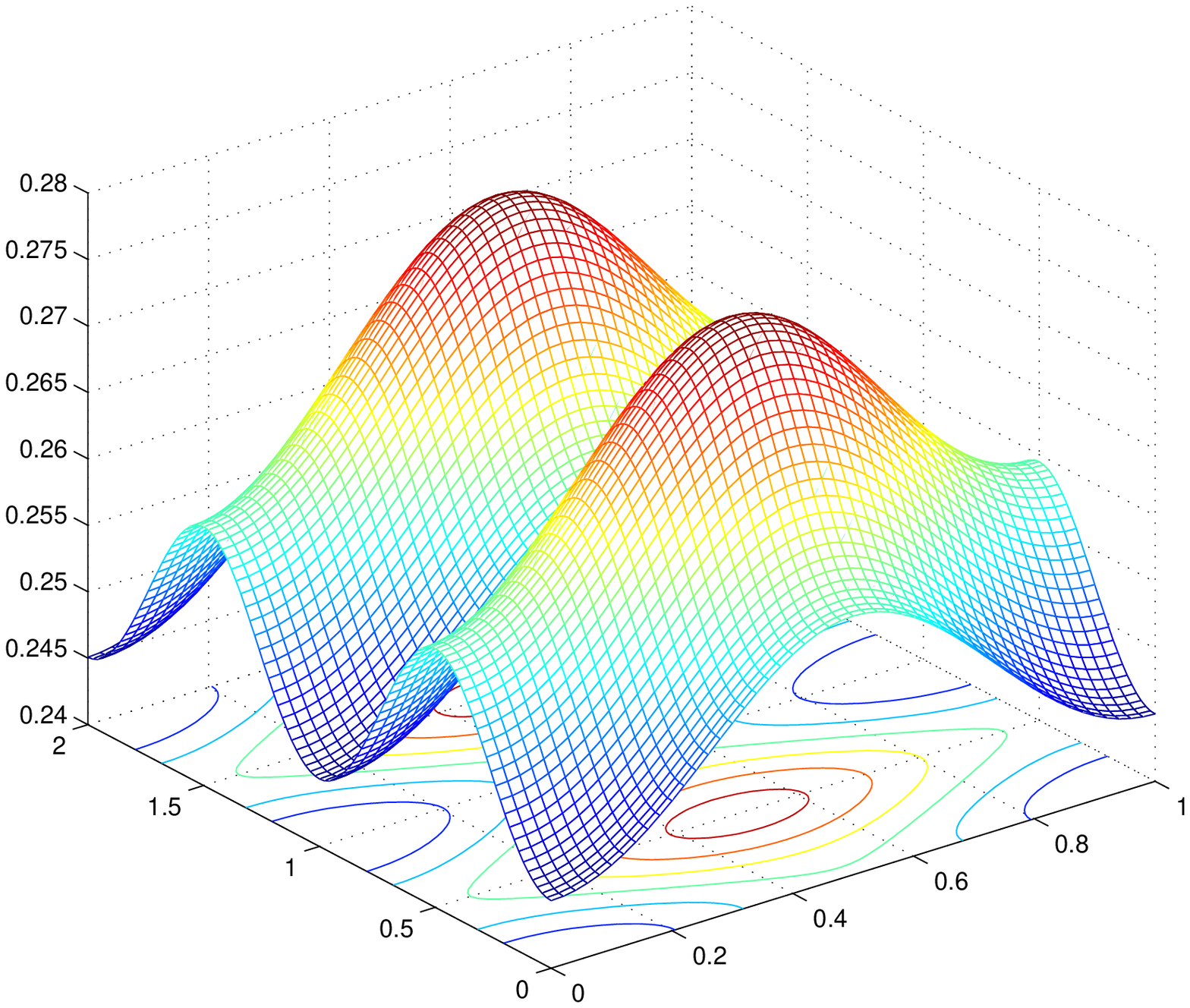}
\caption{The dispersion law $\epsilon(\mathbf k)$ for $l=1$, $N=2$, $K=-1$ and
the geometry of one vortex in a square cell. Enlarged
Brillouin cell corresponding to the real-lattice cell of size $ 2\times 1$ in
units of base periods.}
\label{fig:fig3}
\end{figure*}
\end{center}


\begin{thebibliography}{99}


\bibitem{qh} The Quantum Hall Effect, Ed. by R. Prange and S. M. Girvin
(Springer, New York, 1987; Mir, Moscow, 1989).

\bibitem{nqh} New Perspectives in Quantum Hall Effects, Ed. by S. Das Sarma
and A. Pinczuk (Wiley,  1997).

\bibitem{lgh1} R. B. Laughlin, Phys. Rev. B {\bf22}, 5632 (1981).

\bibitem{lgh2} R. B. Laughlin, Phys. Rev. Lett. {\bf50}, 1395 (1983).

\bibitem{j1} J. K. Jain, Phys. Rev. Lett. {\bf63}, 199 (1989).

\bibitem{j2} J. K. Jain, Phys. Rev. B {\bf41}, 7653 (1990).

\bibitem{h1} B. I. Halperin, P. A. Lee, and N. Read, Phys. Rev. B {\bf47}, 7312
(1993).
\bibitem{tk} V.K.Tkachenko,ZhETF,v{\bf49} 6,p1876 (1965)
\bibitem{ui} E.T.Whitaker,R.N.Watson, A Course of Modern Analysis,pII,
Cambridge At the  University Press (1927)

\bibitem{br} E.Brown,Phys.Rev,v{\bf 133} 4A,A1038(1964)
\bibitem{zk} J.Zak, PR{\bf 134},A1602(1964)
\bibitem{lp9} E. M. Lifshitz and L. P. Pitaevski, Course of Theoretical
Physics, Vol. [IX]: Statistical Physics, Part 2(Nauka, Moscow, 1978;
Pergamon, New York, 1980)
\bibitem{dlg1} V.S,Khrapai,A.A.Shashkin,M.G.Trokina,V.T.Dolgopolov,
V.Pellegrini,F.Beltram,G.Biasol,L.Sorba, PRL {\bf 99},086802,(2007)
\bibitem{LL2}Landau L.D.Lifshits E.M.,Course of Theoretical Physics,Vol.[VIII], Electrodynamics of Continuous Media
Fizmatlit,Moscow,2003.
\bibitem{BIE} Bychkov Yu.A,Iordanskii S.V,Eliashberg G.M. JETP Lett.v33,issue3
(1981)
\bibitem{V} M.M.Salomaa, G.E. Volovik, Rev.Mod.Phys. , vol{\bf 59}, 3, part I, p 533,(1987)
\bibitem{S} S.Sondhi, A.Kahlrede, S.Kivelson, Phys.Rev. {\bf B47}, p 1618 (1993)
\bibitem{BP} G.L. Bir, G.E.Pikus, Symmetry and the deformation effects in semyconductors, Moscow Fizmatlit(1972)
\bibitem{w1} R.L.Willet,M.A.Paalanen,R.R.Ruel,K.W.West,L.N.Pfeiffer,D.J.Bishop
Phys.Rev.Lett.,{\bf 63},112 (1990)
\bibitem{O} E.Onofri, arXiv:0804.3673v[quant-ph],30Apr2008
\bibitem{I} S.V.Iordanski , Pisma v ZhETF,vol{\bf 87}, iss 10.p 669,(2008).The statement on the additional macroscopical degeneracy of the electron energy in this paper is wrong.
\end{thebibliography}
\end{document}